\documentstyle[12pt]{article}
\begin{document}

\title
{Schr\"{o}dinger equations in the constrained space \\
with several initial constraints}
\author{ Takashi Matsunaga, Tadashi Miyazaki, Masanobu Nojiri, \\
\bigskip
Chi\'e Ohzeki and Motowo Yamanobe \\
{\it Department of Physics, Science University of Tokyo,} \\
{\it Kagurazaka, Shinjuku-ku, Tokyo 162, Japan}}
\date{}
\maketitle
\vspace{20mm}

\vspace{20mm}
A general system constrained with {\it several} initial constraint
conditions is quantized based on the Dirac formalism and the
Schr\"{o}dinger equation for this system is obtained.
These constraint conditions are now allowed to depend not only on
the coordinates but also on the velocities.
It is shown that the hermiticity for the observables of the system
restricts the geometrical structure of our world.

\begin{flushleft}
PACS number(s): 03.65.-w, 02.90.+p
\end{flushleft}

\newpage
\section{INTRODUCTION}
\label{sec:intro}

Almost every physical theory,
electromagnetism, gravity and etc.,
is treated as the theory of a constrained system.
The theoretical framework of the constrained system
was invented in 1950's by Dirac with 
the Hamiltonian formalism \cite{dirac}\cite{sundermeyer}.
The essence of the Dirac method is as follows.
Due to the Hamiltonian formalism,
we normally use the canonical {\it Poisson brackets} in the
{\it unconstrained phase space}.
In the {\it constrained phase space}, however, a new type of
brackets -- {\it Dirac brackets} -- must be introduced
{\it \`{a} la} Dirac.
In passing from the classical to the quantum theory,
we replace these Dirac brackets by 
the commutators $(\times 1/i\hbar)$.
To obtain the Schr\"{o}dinger equation,
we have to find the representation of the commutation relations,
{\it i.e.}, we must obtain the representation of the operator 
variables.

The classical theories of constrained systems are completely
described in Hamiltonian formalism
with the Dirac method\footnote{Its details are explored in Appendix}.
However, when we pass to the quantum theory,
we meet some problems\cite{falck}\cite{ishikawa}.
One of such problems is that
Falck and Hirshfeld presented a simple example of the Dirac method 
applied 
to the dynamics of a point particle 
They quantized a system with a particle on a sphere 
in the 3-dimensional Euclidean space.
In this case the Dirac brackets of the variables, expressed in 
the Cartesian coordinates, are {\it not canonical}.
Here ``$\cdots$ canonical'' means that 
the Dirac bracket between a coordinate variable 
and its conjugate momentum variable is 1 and that other Dirac 
brackets with 
respect to the variables vanish.
As is clear from this example,
the Dirac brackets among the conjugate variables are generally not 
canonical,
and the difficulties arise to find out the representation of 
the operator 
variables when we pass to the quantum theory \cite{ishikawa}.

Homma {\it et al.} suggested a new method to overcome these 
difficulties for a constrained particle system on a sphere and on 
a general hypersurface \cite{homma1}\cite{homma2}.
A constraint $f(q^i)=0$ with $q^i$, the coordinates of a particle, 
is 
evidently equivalent to $df/dt=0$ with a vanishing constraint 
{\it at some instant}.
They defined the system under the constraint $f(q^i)=0$ as system(I) 
and that under the constraint $df/dt=0$ as system(II).
The two systems should be equivalent both in the classical mechanics 
and in the quantum mechanics.
The Dirac brackets in system(I) are {\it not} canonical, but those 
in system(II) are canonical, whichever is the reason why they 
introduced system(II).
They, therefore, easily found the representation of the operators 
in the latter system.
Now let us call these constraint conditions introduced initially to 
make 
a particle constrained on a specific manifold, by the name 
``{\it initial} constraint conditions''.

Homma {\it et al.} dealt with the system constrained only with 
a {\it single} 
initial constraint condition, {\it i.e.},  the system constrained 
on a hypersurface.
In other words, they formulated the particle dynamics 
in the constrained coordinate space whose dimension is only reduced 
by one from that of the unconstrained coordinate space: 
in the constrained 
phase space the dimension is reduced by two from that of the 
unconstrained phase space.
However, the system with several initial constraint conditions has 
been discarded up to now.

Following their method, we, in this paper, treat {\it several} 
initial 
constraint conditions to complete the dynamics of a particle on 
various manifolds.
Moreover these conditions are made to depend not only on 
the coordinates but also {\it on the velocities}.
Thus if we succeed to construct the theory considering the several 
initial 
constraint conditions depending on coordinate and velocity variables,
we are to have the {\it general} theory of constrained systems with 
arbitrary 
initial constraint conditions. 

This paper is composed as follows.
In Sec.~\ref{sec:2}, we shall review the method of 
Homma {\it et al.},
restricted to an example of a particle constrained on a sphere of 
arbitrary 
dimensions.
In Sec.~\ref{sec:3}, we shall deal with several initial constraint 
conditions 
depending only on the coordinates, and in Sec.~\ref{sec:4}, with 
these 
depending both on the coordinates and the velocities of a particle.
The last section will be devoted to the summary and discussions.

In present paper, the summation convention, 
according to which summation over the repeated indices is meant, is 
adopted.

\section{QUANTIZATION ON A SPHERE}
\label{sec:2}

Now, we consider, as system(I), a nonrelativistic particle 
(mass : $m$) 
constrained on an $(N-1)$-dimensional sphere 
(radius : $\sqrt{A}$ and $N\geq2$) in the $N$-dimensional Euclidean 
space.
The initial constraint condition in this system is 
$f(q)\equiv q^iq_i-A=0$, and the Lagrangian is
\begin{equation}
L_{\rm I}=\frac{1}{2}m\dot{q}^i\dot{q}_i+\lambda(q^i q_i-A)\;\;\;\;\;
(i=1,2,\cdots,N) \;,
\label{2.1}
\end{equation}
where \[
\dot{q}^i\equiv\frac{dq^i}{dt}\;,
\]
the $q^i$ denote the coordinates of the particle and $\lambda$ is 
a Lagrange multiplier.
The conjugate momenta are defined as 
\begin{eqnarray}
p_i&\equiv&\frac{\partial L_{\rm I}}{\partial \dot{q}^i}=m\dot{q}^i
\;, 
\label{2-2} \\
p_\lambda&\equiv&\frac{\partial L_{\rm I}}{\partial \dot{\lambda}}=0
\;.
\label{2-3}
\end{eqnarray}
Equation (\ref{2-3}) is a primary constraint condition
\begin{eqnarray}
\phi_1 \equiv p_\lambda\approx 0\;,
\label{2-4}
\end{eqnarray}
where `` $\approx$ '' represents Dirac's weak equality.
Following the Dirac 
method we obtain the secondary constraint conditions
\begin{eqnarray}
\phi_2 &\equiv& q^iq_i-A\approx 0 \;, 
\label{2.2} \\
\phi_3 &\equiv& q^i p_i\approx 0 \;,
\label{2.3} \\
\phi_4 &\equiv& p^ip_i+2m\lambda q^iq_i\approx 0 \;.
\label{2-7}
\end{eqnarray}
By Eq.(\ref{2-7}), 
the variable $\lambda$ is written in terms of $q^i$ and $p_i$:
\begin{eqnarray}
\lambda=\frac{p^ip_i}{2mq^iq_i}\;.
\label{2-8}
\end{eqnarray}
We, further, eliminate $p_\lambda$ and $\lambda$ from this system 
with Eqs.(\ref{2-4}) and (\ref{2-8}), respectively.
Therefore the variables of this system are $q^i$ and $p_i$, 
and the remaining constraint conditions are $\phi_2 \approx 0$
and $\phi_3 \approx 0$. 
The Dirac brackets of the variables are calculated as follows:
\begin{eqnarray}
\{q^i,q^j\}_{\rm D}&=&0\;,
\label{2.4}   \\
\{q^i,p_j\}_{\rm D}&=&\delta^i\!_j-\frac{q^i q_j}{A}\;,
\label{2.5}   \\
\{p_i,p_j\}_{\rm D}&=&\frac{p_i q_j-p_j q_i}{A} \;.
\label{2.6}
\end{eqnarray}
We find that the Dirac brackets (\ref{2.4})--(\ref{2.6}) are not 
canonical.
After we have introduced the Dirac brackets, the extended 
Hamiltonian is
\begin{equation}
H_{\rm I E}=\frac{1}{2m}p^i p_i \;.
\label{2.7}
\end{equation}

In the canonical quantization method,
we replace the Dirac brackets (\ref{2.4})--(\ref{2.6}) by
the commutators ($\times 1/i\hbar$) and regard the variables as 
operators: 
\begin{eqnarray}
\left[\hat{q}^i,\hat{q}^j\right] &=& 0 \;,
\label{2.11}   \\
\left[\hat{q}^i,\hat{p}_j\right] &=& i\hbar\left(
\delta^i\!_j-\frac{\hat{q}^i\hat{q}_j}{A}\right) \;,
\label{2.12} \\
\left[\hat{p}_i,\hat{p}_j^{}\right] &=& i\hbar
\frac{\hat{p}_i\hat{q}_j-\hat{p}_j\hat{q}_i}{A} \;,
\label{2.13} 
\end{eqnarray}
where the symbol with caret `` $\hat{}$ '' designates an operator.

In order to obtain the Schr\"{o}dinger equation, 
we must find the coordinate representation of $\hat{p}_i$.
In our simple case, it is possible to obtain its representation 
by the trial-and-error method.
However, since the commutation relations of $q^i$ and $p_i$ are 
complicated
in general cases, 
it is difficult to do that.
Therefore we discuss the quantization method {\it \`{a} la} 
Homma {\it et al.}

Firstly, we consider the constrained system 
with the initial constraint condition $q^i\dot{q}_i=0$ and refer it 
to system(II).
This condition is the time derivative of 
the initial constraint condition in system(I): $\dot{f}(q)=2 q^i
\dot{q}_i=0$.
The Lagrangian of system(II) is given by
\begin{eqnarray}
L_{\rm II}=\frac{1}{2}m\dot{q}^i\dot{q}_i+\lambda(q^i\dot{q}_i)\;.
\label{2.14}
\end{eqnarray}
The conjugate momenta are
\begin{eqnarray}
\Pi_i&\equiv&\frac{\partial L_{\rm II}}
{\partial \dot{q}^i}=m\dot{q}_i+\lambda q_i\;,
\label{2-17} \\
\Pi_\lambda&\equiv&\frac{\partial L_{\rm II}}{\partial \dot{\lambda}}
=0\;.
\label{2-18}
\end{eqnarray}
The primary constraint condition is
\begin{eqnarray}
\chi_1\equiv\Pi_\lambda\approx 0\;,
\label{2-19}
\end{eqnarray}
and the secondary constraint condition is 
\begin{eqnarray}
\chi_2\equiv q^i\Pi_i-\lambda q^iq_i\approx 0\;.
\label{2-20}
\end{eqnarray}
In system(II),
the Dirac brackets of $q^i$ and $\Pi_i$ are
\begin{eqnarray}
\{q^i,q^j\}_{\rm D} &=& 0 \;,
\label{2.15}  \\
\{q^i,\Pi_j\}_{\rm D} &=& \delta^i\!_j \;,
\label{2.16} \\
\{\Pi_i,\Pi_j\}_{\rm D} &=& 0 \;.
\label{2.17}
\end{eqnarray}
Equations (\ref{2.15})--(\ref{2.17}) show that
the Dirac brackets are {\it canonical} in system(II).
The extended Hamiltonian is
\begin{eqnarray}
H_{\rm II E} &=& \frac{1}{2m}
			\left(
			\Pi_i-\frac{q^iq_j}{q^2}\Pi_j
			\right)^2
\nonumber \\
 &=& \frac{1}{2m}(F^j\!_i\Pi_j)^2 \;,
\label{2.18} 
\end{eqnarray}
where
\begin{eqnarray}
F^j\!_i &\equiv& \delta^j\!_i-\frac{q^j q_i}{q^2} \;,
\label{2.19} \\
q^2&\equiv&q^iq_i\;. \nonumber
\end{eqnarray}
From Eq.(\ref{2.19}), we find that $F^j\!_i$ is the projection 
matrix which projects an arbitrary vector onto the sphere.

In the above discussion,
we treat $(q^i,\Pi_i,\lambda,\Pi_\lambda ; i=1,2,...,N)$
as the variables of system(II).
However we can reduce them to $(q^i,\Pi_i ; i=1,2,...,N)$ by
solving the equations $\chi_1\approx 0$ and
$\chi_2 \approx 0$ to represent $\Pi_\lambda$ and $\lambda$ in the 
other 
variables.
In this case, 
since all constraint conditions are disappeared, 
we can deal with the system through the Poisson brackets
constructed with $(q^i,\Pi_i ; i=1,2,...,N)$.
The Poisson brackets of $q^i$ and $\Pi_i$ 
take the same form as in Eqs.(\ref{2.15})--(\ref{2.17}).

Secondly, we introduce new momentum variables $P_i$:
\begin{eqnarray}
P_i &\equiv& F^j\!_i\Pi_j \;.
\label{2.20}
\end{eqnarray}
which are interpreted as projected momentum variables. 
The Dirac brackets including $P_i$ are calculated as follows:
\begin{eqnarray}
\{q^i,P_j\}_{\rm D} &=& \delta^i\!_j-\frac{q^iq_j}{q^2}=F^i\!_j \;,
\label{2.21}  \\
\{P_i,P_j\}_{\rm D} &=& \frac{P_iq_j-P_jq_i}{q^2} \;.
\label{2.22}
\end{eqnarray}
Investigate the time development of $q^2$,
\begin{equation}
\frac{dq^2}{dt} = \{q^2,H_{\rm II E}\}_{\rm D} = 0 \;,
\label{2.23}
\end{equation}
and we find that $q^2$ is a constant of motion.
Putting $q^2= A$ at $t=0$ as the initial condition of 
the differential equation (\ref{2.23}),
the Dirac brackets (\ref{2.21}) and (\ref{2.22}) take the same form 
as in 
Eqs.(\ref{2.5}) and (\ref{2.6}).
We further obtain, from Eq.(\ref{2.20}), the constraint condition 
whose form 
is the same as that of Eq.(\ref{2.3}):
\begin{eqnarray}
q^iP_i=q^iF^j\!_i\Pi_j=0\;.
\label{2-28}
\end{eqnarray}

Finally, 
we regard $P_i$ as the momenta $p_i$ of system(I).
In this way the constraint structure of 
system(II) is equivalent to that of system(I).
Therefore system(I) and system(II) have the same content 
in the classical theory. 

In order to quantize system(II),
we replace the Dirac brackets (\ref{2.15})--(\ref{2.17}) by 
the commutators ($\times 1/i\hbar$) of the operators:
\begin{eqnarray}
\left[\hat{q}^i,\hat{q}^j\right] &=& 0 \;,
\label{2.24} \\
\left[\hat{q}^i,\hat{\Pi}_j\right] &=& i\hbar\delta^i\!_j \;,
\label{2.25} \\
\left[\hat{\Pi}_i,\hat{\Pi}_j\right] &=& 0 \;.
\label{2.26}
\end{eqnarray}
From these commutation relations, we immediately have the coordinate 
representation of $\hat{\Pi}_i$:
\begin{eqnarray}
\hat{\Pi}_i=-i\hbar\frac{\partial}{\partial{q}^i} \;.
\label{2.27}
\end{eqnarray}
Therefore it is easy to obtain the Schr\"odinger equation of 
system(II) in the 
coordinate representation.

Requiring the equivalence between system(I) and system(II) in
the quantum theory and making use of Eq.(\ref{2.27}),
we obtain the Schr\"odinger equation for system(I). 

To sum up, in the quantum case, we adopt $\hat{q}^2=A$ 
as the initial condition,
and then regard $\hat{P}_i$ as being equal to $\hat{p}_i$ after 
symmetrizing the order of operators in Eq.(\ref{2.20}):
\begin{eqnarray}
\hat{p}_i=\frac{1}{2}
  \left[
    \hat{F}^j\!_i\hat{\Pi}_j+\hat{\Pi}_j\hat{F}^j\!_i
  \right] \;.
\label{2.28}
\end{eqnarray}
We, of course, have all constraint conditions of system(I) based 
on those of 
system(II) in quantum level.
Substitution of Eq.(\ref{2.28}) into the quantum Hamiltonian 
of system(I)
\begin{eqnarray}
\hat{H}_{\rm I\,E}=\frac{1}{2m}\hat{p}^i\hat{p}_i \;,
\label{2.29}
\end{eqnarray}
gives us the Hamiltonian of the system(II).
The resultant quantum Hamiltonian of system(II) is reduced to,
with an appropriate ordering,
\begin{eqnarray}
\hat{H}_{\rm II\,E}=\frac{1}{8m}
\left(
  \hat{F}^j\!_i\hat{\Pi}_j
  +\hat{\Pi}_j\hat{F}^j\!_i
\right)^2 \;.
\label{2.30}
\end{eqnarray}
System(II) thus becomes equivalent to system(I) also in quantum 
level.
In order to obtain the coordinate representation of the Hamilton
operator of 
system(I),
we substitute Eq.(\ref{2.27}) into Eq.(\ref{2.30}).

\section{SEVERAL CONSTRAINT CONDITIONS DEPENDING ONLY ON 
COORDINATE VARIABLES}
\label{sec:3}

In this section,
we treat the dynamics of a nonrelativistic particle of mass $m$ 
in an $N$-dimensional Riemannian space with a Riemannian metric 
$g_{ij}(q)$.
Let a particle be constrained with $M$ initial constraint conditions
\begin{equation}
f_s(q)=0\;\;\;\;\;(s=1,2,\cdots,M;M<N) \;,
\label{3.1}
\end{equation}
where $s$ is the index to distinguish various initial constraint 
conditions.
Namely, the particle is constrained in an $(N-M)$-dimensional 
subspace.
Here we assume that the constraint conditions depend only on the 
coordinate 
variables $q^i$ and that these conditions are differentiable with 
respect to 
$q^i$.
The Lagrangian $L_{\rm I}$ in this system is expressed as follows:
\begin{equation}
L_{\rm I}=\frac{1}{2}mg_{ij}\dot{q}^i\dot{q}^j+A_i(q)\dot{q}^i-V(q)
+\lambda^s f_s(q) \;\;\;\;\;(i,j=1,2,\cdots,N) \;,
\label{3.2}
\end{equation}
where\[
\dot{q}^i\equiv\frac{dq^i}{dt} \;,
\]
and $V(q)$, $A_i(q)$ and $\lambda^s$ are the scalar potential, the 
vector 
potentials and the Lagrange multipliers, respectively.
The momenta conjugate to $q^i$ and $\lambda^s$ are
\begin{eqnarray}
p_i&\equiv&\frac{\partial L_{\rm I}}{\partial \dot{q}^i}
=mg_{ij}\dot{q}^j+A_i(q) \;,
\label{3.3} \\
p'_s&\equiv&\frac{\partial L_{\rm I}}{\partial \dot{\lambda}^s}=0 \;.
\label{3.4}
\end{eqnarray}
The primary constraint conditions are
\begin{equation}
\phi_{1s}\equiv p'_s\approx 0 \;.
\label{3.5}
\end{equation}
We thus have the total Hamiltonian
\begin{equation}
H_{\rm IT}=\frac{1}{2m}g^{ij}(p_i-A_i)(p_j-A_j)+V(q)-\lambda^s f_s(q)
+v^sp'_s   \;,
\label{3.6}
\end{equation}
where\[
v^s\equiv\dot{\lambda}^s \;,
\]
and $v^s$ are the multipliers for Eq.(\ref{3.5}).
For the consistency that a constraint condition should hold also 
after the 
time development, we succeedingly have secondary constraint 
conditions
\begin{eqnarray}
\phi_{2s}&\equiv& f_s(q)\approx0 \;,
\label{3.7}   \\
\phi_{3s}&\equiv&\frac{1}{m}g^{ij}\frac{\partial{f}_s}{\partial{q}^i}
(p_j-A_j)
\approx0 \;,  \label{3.8} \\
\phi_{4s}&\equiv&\frac{1}{m^2}\left[g^{kl}\left\{\frac{\partial}
{\partial{q}^k}
\left( g^{ij}\frac{\partial{f}_s}{\partial{q}^i}\right)p_j
-\frac{\partial}{\partial{q}^k}\left(g^{ij}\frac{\partial{f}_s}
{\partial{q}^i}A_j\right)\right\}(p_l-A_l)\right. \nonumber \\
& &\left.-\frac{1}{2}
\frac{\partial{g}^{kl}}{\partial{q}^i}g^{ij}\frac{\partial{f}_s}
{\partial{q}^j}
p_k p_l+\frac{\partial}{\partial{q}^j}(g^{kl}A_l)g^{ij}
\frac{\partial{f}_s}{\partial{q}^i}p_k
-\frac{1}{2}g^{ij}
\frac{\partial{f}_s}{\partial{q}^i}\frac{\partial}{\partial{q}^j}
(g^{kl}A_k A_l)\right] \nonumber \\ 
& &-\frac{1}{m}g^{ij}\frac{\partial{f}_s}{\partial{q}^i}
\frac{\partial{V}}
{\partial{q}^j}+\frac{1}{m}D_{st}\lambda^t\approx0 \;,
\label{3.9}
\end{eqnarray}
where\[
D_{st}\equiv g^{ij}\frac{\partial{f}_s}{\partial{q}^i}
\frac{\partial{f}_t}
{\partial{q}^j} \;.
\]
Since the initial constraint conditions $f_s=0$ are independent, 
{\it i.e.},
the hypersurfaces $f_s=0$ are independent each other, the normals to 
the surfaces ${\partial f_s(q)}/{\partial q^i}$ are linearly 
independent.
This means that the determinant of $(D_{st})$ does not vanish, and 
$\lambda^s$ are rewritten as the functions of $q^i$ and $p_i$ with 
Eq.(\ref{3.9}).
Moreover $v^s$ are determined as the functions of $q^i$ and $p_i$ 
for 
the consistency of $\phi_{4s}$.
Hence all the physical quantities are expressed with $q^i$ and $p_i$.
We calculate the Poisson brackets among $\phi_{\alpha s}\;
(\alpha=1,2,3,4)$:
\begin{eqnarray}
\{\phi_{1s},\phi_{4t}\}&=&-\frac{1}{m}D_{st} \;,
\label{3.10} \\
\{\phi_{2s},\phi_{3t}\}&=&\frac{1}{m}D_{st} \;,
\label{3.11}
\end{eqnarray}
and all other Poisson brackets vanish.
Hence all $\phi_{\alpha s}=0$ are second class constraint conditions.
The Dirac brackets among $q^i$ and $p_i$ are
\begin{eqnarray}
\{q^i,q^j\}_{\rm D}&=&0 \;,
\label{3.12}   \\
\{q^i,p_j\}_{\rm D}&=&\delta^i\!_j-g^{ki}\left( D^{-1}\right)^{st}
\frac{\partial{f}_s}
{\partial{q}^k}\left.\frac{\partial{f}_t}
{\partial{q}^j}\right|_{f=0} \;,
\label{3.13}    \\
\{p_i,p_j\}_{\rm D}&=&-\frac{\partial{f}_s}{\partial{q}^i}
(D^{-1})^{st}
\left.\left[\frac{\partial}{\partial{q}^j}
\left(g^{kl}\frac{\partial{f}_t}{\partial{q}^k}\right)p_l
-\frac{\partial}{\partial{q}^j}
\left(g^{kl}\frac{\partial{f}_t}{\partial{q}^k}A_l\right)
\right]\right|_{f=0}-(i\leftrightarrow j) ,
\label{3.14}
\end{eqnarray}
where ${}|_{f=0}$ denotes that we impose all initial constraint 
conditions 
$f_s=0$ after having calculated the Poisson brackets.
We exchange the indices $i$ and $j$ in the first term and the term 
thus 
made is designated by $(i\leftrightarrow j)$.
With the Dirac brackets, the extended Hamiltonian is
\begin{equation}
H_{\rm IE}=\frac{1}{2m}g^{ij}(p_i-A_i)(p_j-A_j)+V(q) \;.
\label{3.15}
\end{equation}

We now consider system(II) following the procedure in 
Sec.~\ref{sec:2}.
In system(II), the Lagrangian $L_{\rm II}$ is
\begin{equation}
L_{\rm II}=\frac{1}{2}mg_{ij}\dot{q}^i \dot{q}^j+A_i(q)\dot{q}^i-V(q)
+\lambda^s\dot{f}_s(q) \;\;\;\;\;(i,j=1,2,\cdots,N)\;.
\label{3.16}
\end{equation}
The momenta conjugate to $q^i$ and $\lambda^s$ are
\begin{eqnarray}
\Pi_i&\equiv&\frac{\partial L_{\rm II}}{\partial \dot{q}^i}
=mg_{ij}\dot{q}^j+A_i+\lambda^s\frac{\partial f_s}{\partial q^i}\;,
\label{3.17} \\
\Pi'_s&\equiv&\frac{\partial L_{\rm II}}{\partial \dot{\lambda}^s}=0 
\;.
\label{3.18}
\end{eqnarray}
Therefore the primary constraint conditions are
\begin{equation}
\chi_{1s}\equiv\Pi'_s\approx0 \;.
\label{3.19}
\end{equation}
Thus the total Hamiltonian is
\begin{equation}
H_{\rm IIT}=\frac{1}{2m}g^{ij}\left(\Pi_i-A_i-\lambda^s
\frac{\partial f_s}{\partial q^i}\right)\left(\Pi_j-A_j-\lambda^t
\frac{\partial f_t}{\partial q^j}\right)+V(q)+u^s\Pi'_s \;,
\label{3.20}
\end{equation}
where\[
u^s\equiv\dot{\lambda}^s \;,
\]
and $u^s$ are the multipliers for Eq.(\ref{3.19}).
The consistency requires the following secondary constraint 
conditions:
\begin{equation}
\chi_{2s}\equiv\frac{1}{m}g^{ij}\frac{\partial f_s}{\partial q^i}
\left(\Pi_j-A_j-\lambda^t\frac{\partial f_t}{\partial q^j}\right)
\approx0 \;.
\label{3.21}
\end{equation}
Eliminating $\lambda^s$ and $u^s$ in the same way as in system(I), 
all the physical quantities are expressed with $q^i$ and $\Pi_i$.
We calculate the Poisson brackets among $\chi_{\alpha s}\;
(\alpha=1,2)$:
\begin{equation}
\{\chi_{1s},\chi_{2t}\}=\frac{1}{m}D_{st} \;,
\label{3.22}
\end{equation}
and all other Poisson brackets vanish,
indicating that all $\chi_{\alpha s}=0$ are second class constraint 
conditions.
The Dirac brackets among $q^i$ and $\Pi_i$ are
\begin{eqnarray}
\{q^i,q^j\}_{\rm D}&=&0 \;,
\label{3.23}   \\
\{q^i,\Pi_j\}_{\rm D}&=&\delta^i\!_j \;,
\label{3.24}   \\
\{\Pi_i,\Pi_j\}_{\rm D}&=&0 \;.
\label{3.25}
\end{eqnarray}
The extended Hamiltonian is
\begin{equation}
H_{\rm IIE}=\frac{1}{2m}g^{ij}F^k\!_i(\Pi_k-A_k)F^l\!_j(\Pi_l-A_l)
+V(q) \;,
\label{3.26}
\end{equation}
where
\begin{equation}
F^k\!_i\equiv \delta^k\!_i
-g^{kl}\left(D^{-1} \right)^{st}\frac{\partial f_s}
{\partial q^l}\left.\frac{\partial f_t}{\partial q^i}\right|_{f=0} 
\;.
\label{3.27}
\end{equation}
Here we define new momentum variables $P_i$:
\begin{equation}
P_i-A_i\equiv F^k\!_i(\Pi_k-A_k) \;.
\label{3.28-1}
\end{equation}
The Dirac brackets including $P_i$ are calculated as follows:
\begin{equation}
\{q^i,P_j\}_{\rm D}=F^i\!_j \;,
\label{3.28}   
\end{equation}
\begin{equation}
\{P_i,P_j\}_{\rm D}=-\frac{\partial{f}_s}{\partial{q}^i}(D^{-1})^{st}
\left.\left[\frac{\partial}{\partial{q}^j}
\left(g^{kl}\frac{\partial{f}_t}{\partial{q}^k}\right)P_l
-\frac{\partial}{\partial{q}^j}
\left(g^{kl}\frac{\partial{f}_t}{\partial{q}^k}A_l\right)
\right]\right|_{f=0}-(i\leftrightarrow j) \;.
\label{3.29}
\end{equation}
Moreover we have the time development of $f_s$
\begin{equation}
\frac{df_s}{dt}=\{f_s,H_{\rm IIE}\}_{\rm D}=0 \;,
\label{3.30}
\end{equation}
which shows that $f_s$ are the constants of motion.
When we choose $f_s=0$ as the initial conditions of the differential 
equations 
(\ref{3.30}) ({\it i.e.}, at $t=0$), the Dirac brackets (\ref{3.28}) 
and 
(\ref{3.29}) take the same forms as in Eqs.(\ref{3.13}) and 
(\ref{3.14}).
Note that since $(P_i-A_i)$ are the tangential components of 
$(\Pi_k-A_k)$ 
to the surfaces and $\partial f_s/\partial q^i$ are the normal 
components, $(P_i-A_i)$ just satisfy the constraint conditions 
(\ref{3.8}) 
with $p_i$, replaced by $P_i$. 
The argument mentioned above means that the constraint 
structure of system(I) is equivalent to that of system(II).
We take $f_s=0$ as the initial conditions and identify $P_i$ with 
the momenta 
$p_i$ in system(I).

We pass to the quantum theory under the requirement of the above 
equivalence of both systems(I) and (II).
For the hermiticity, we need to symmetrize the order of the products 
of operators expressing observables.
Firstly, we symmetrize the momentum operators:
\begin{equation}
\hat{p_i}-\hat{A_i}=\frac{1}{2}[\hat{F}^k\!_i(\hat{f}=0),\hat{\Pi}_k
-\hat{A}_k]
_{+} \;,
\label{3.31}
\end{equation}
where $[\,\,\,,\,\,\,]_{+}$ denotes the anticommutator.
The commutation relations including $\hat{\Pi}_i$ are
\begin{eqnarray}
\left[\hat{q}^i,\hat{\Pi}_j\right]&=&i\hbar \delta^i\!_j \;,
\label{3.32} \\
\left[\hat{\Pi}_i , \hat{\Pi}_j\right]&=&0 \;.
\label{3.33}
\end{eqnarray}
These are canonical.
The coordinate representation of $\hat{\Pi}_i$ is
\begin{equation}
\hat{\Pi}_i=-i\hbar\frac{\partial}{\partial {q}^i}
-\frac{1}{2}i\hbar\left\{
\begin{array}{c}
k \\
ki
\end{array}
\right\}(q) \;.
\label{3.34}
\end{equation}
Here we use the Christoffel symbol
\begin{equation}
\left\{
\begin{array}{c}
j \\
ki
\end{array}
\right\}(q)\equiv \frac{1}{2}g^{jn}
(\partial_k g_{in}+\partial_i g_{kn}
-\partial_n g_{ki}) \;.
\label{3.35}
\end{equation}

Secondly, we obtain the quantum Hamiltonian by symmetrization.
We write down the Hamiltonian, based on Eqs.(\ref{3.15}) and 
(\ref{3.31}), as follows \cite{dewitt}:
\begin{equation}
\hat{H}_{\rm IIE}=\frac{1}{2m}[\hat{F}^k\!_i(\hat{f}=0),\hat{\Pi}_k
-\hat{A}_k]_
{+}\hat{g}^{ij}
[\hat{F}^l\!_j(\hat{f}=0),\hat{\Pi}_l-\hat{A}_l]_{+}+V(\hat{q})
+{\hbar}^2\hat{Q}\;,
\label{3.36}
\end{equation}
where
\begin{eqnarray}
\hat{Q} &\equiv& \frac{1}{2m}
  \left[
    F^k\!_iF^l\!_j
      \left(
        g^{ij}
	\left\{
	\begin{array}{c}
	a \\
	a\,k
	\end{array}
	\right\}
        -g^{jb}
	\left\{
	\begin{array}{c}
	i \\
	k\,b
	\end{array}
	\right\}
      \right) 
    \partial_l
    +F^l\!_i g^{ij}
      \left(
	\partial_kF^k\!_j
      \right)
    \partial_l
\right. \nonumber \\
& & \left.
    +\frac{1}{2}F^k\!_i\partial_k
      \left\{
	g^{ij}
	  \left(
	    F^l\!_j
	    \left\{
	    \begin{array}{c}
	    a \\
	    a\,l
	    \end{array}
	    \right\}
	    +\partial_lF^l\!_j
	  \right)
      \right\}
\right. \nonumber \\
& & \left.
    +\frac{1}{4}
      \left(
	F^k\!_i
	\left\{
	\begin{array}{c}
	b \\
	b\,k
	\end{array}
	\right\}
        +\partial_kF^k\!_i
      \right)
    g^{ij}
      \left(
	F^l\!_j
	\left\{
	\begin{array}{c}
	a \\
	a\,l
	\end{array}
	\right\}
      +\partial_lF^l\!_j
    \right)
  \right] \;.
\label{3.37}
\end{eqnarray}
The quantum mechanical potential $\hat{Q}$, typical of quantum 
mechanics, is 
a term which makes the Schr\"{o}dinger equation be covariant 
under general coordinate transformations.
The hermiticity of the Hamiltonian (\ref{3.36}) leads to the 
vanishing of 
the antisymmetric parts for those multiplied by 
the metric $g^{ij}$.
Hence we have the hermiticity conditions
\begin{eqnarray}
-F^k\!_i\partial_k F^l\!_j+(\partial_iF^h\!_j)F^l\!_h
+F^k\!_j\partial_kF^l\!_i
-(\partial_j F^h\!_i)F^l\!_h&=&0 \;,
\label{3.38}    \\
\partial_iF^h\!_j\partial_kF^k\!_h
-\partial_jF^h\!_i\partial_kF^k\!_h&=&0 \;,
\label{3.39}  \\
(\delta^k\!_i-F^k\!_i)\partial_k\{(\delta^l\!_j-F^l\!_j)A_l(q)\}
-(i\leftrightarrow j)&=&0 \;,
\label{3.40}    \\
F^l\!_ig^{ij}(\partial_kF^k\!_j)&=&0 \;,
\label{3.41}
\end{eqnarray}
Note that Eq.(\ref{3.41}) comes from the hermiticity of the quantum 
mechanical 
potential.
With these conditions, we finally obtain the Schr\"{o}dinger equation
\begin{eqnarray}
i\hbar\frac{\partial{\Psi}}{\partial{t}}&=&\hat{H}_{\rm E}\Psi 
\nonumber \\
&=&-\frac{{\hbar}^2}{2m}g^{ij}F^k\!_i(F^l\!_j\Psi_{.l})_{.k}
+\frac{i\hbar}{m}F^k\!_iA_kg^{ij}F^l\!_j\Psi_{.l}
+\frac{i\hbar}{2m}(F^k\!_iA_kg^{ij}F^l\!_j)_{.l}\Psi \nonumber \\
&&+\frac{1}{2m}(F^k\!_iA_kg^{ij}F^l\!_jA_l)\Psi+V(q)\Psi \;,
\label{3.42}
\end{eqnarray}
where $\Psi$ is a wave function, and $.k$ denotes the 
{\it covariant derivative} 
with respect to $q^k$, and its concrete expression for $T^j\!_i$ is
\begin{equation}
T^j\!_{i.k}\equiv \nabla_k T^j\!_i=\partial_k T^j\!_i+
	\left\{
	\begin{array}{c}
	j \\
	l\,k
	\end{array}
	\right\} T^l\!_i-
	\left\{
	\begin{array}{c}
	l \\
	i\,k
	\end{array}
	\right\} T^j\!_l \;.
\label{3.43}
\end{equation}

\section{SEVERAL CONSTRAINT CONDITIONS DEPENDING ON
 COORDINATES AND VELOCITIES}
 \label{sec:4}

In this section we consider the system of a particle constrained 
with several initial constraint conditions,
depending not only on 
the coordinates $q^{i}$ but also on the velocities $\dot{q}^i 
(i=1,2,\cdots,N).$

We proceed by taking an example, namely by putting concrete forms 
for 
the initial constraints $f_{s}(\dot{q},q)$.
Other forms for $f_{s}(\dot{q},q)$ are, of course, allowed. 
For instance, $f_{s}(\dot{q},q)$ may include the terms 
whose powers of velocities are greater than one: 
$\alpha'_{sijk\cdots}(q)\dot{q}^{i}\dot{q}^{j}\dot{q}^{k}\cdots$.
But this case leads to a complicated nonlinear equation of motion.
We thus start with a simple (and general) example:
\begin{eqnarray}
f_s(\dot{q},q)=\alpha_{si}(q)\dot{q}^i +\beta_{s}(q)=0\;\;\;\;\;\;
(s=1,2,\cdots,M;M<N)\;,
\label{4.1}
\end{eqnarray}
where both $\alpha_{si}(q)$ and $\beta_s(q)$ are 
the functions of the coordinates $q^{i}$. The conditions (\ref{4.1}) 
are 
assumed to be independent with respect to the velocity variables.
We thus immediately have ${\rm det}(E_{st}) \neq 0$ with 
$E_{st}=g^{ij} \alpha_{si} \alpha_{tj}$. (See the discussion leading 
to 
${\rm det}(D_{st}) \neq 0$ in the previous section.)

The Lagrangian of this system is given by
\begin{equation}
L=\frac{1}{2}mg_{ij}\dot{q}^i\dot{q}^j+A_i(q)\dot{q}^i-V(q)
+\lambda^sf_s(\dot{q},q)
\;\;\;\;\;(s=1,2,...,M)\;,
\label{4.2}
\end{equation}
with $\lambda^{s}$, the Lagrange multipliers.
The momenta $p_i$ conjugate to $q^i$ and $\Pi_s$ conjugate to 
$\lambda^s$ are
\begin{eqnarray}
p_i&=&mg_{ij}\dot{q}^j+A_i(q)+\lambda^s\alpha_{si} \;,
\label{4.3} \\
\Pi_{s}&=&0 \;.
\label{4.4}
\end{eqnarray}
From Eq.(\ref{4.4}), the primary constraints are deduced:
\begin{equation}
\chi_{1s}\equiv\Pi_s\approx 0 \;\;\;\;\; (s=1,2,...,M)\;.
\label{4.5}
\end{equation}
Therefore the total Hamiltonian of the system is given by 
\begin{equation}
H_{\rm T}=\frac{1}{2m}g^{ij}(p_i-A_i-\lambda^s\alpha_{si})
(p_j-A_j-\lambda^t\alpha_{tj})-\lambda^s\beta_s+V(q)+v^s\Pi_s \;,
\label{4.6}
\end{equation}
where $v^s(\equiv\dot{\lambda}^s)$ are 
the Lagrange multipliers for the primary constraints (\ref{4.5}). 

The consistency for the primary constraints leads us to the 
secondary 
constraints
\begin{equation}
\chi_{2s}\equiv-\frac{1}{m}E_{st}\lambda^t+\frac{1}{m}g^{ij}(p_i-A_i)
\alpha_{sj}+\beta_s\approx0 \;.
\label{4.7}
\end{equation}
We can write $\lambda^{s}$ in terms of $q^{i}$ and $p_{i}$ 
as follows:
\begin{equation}
\lambda^s\approx(E^{-1})^{st}[g^{ij}(p_i-A_i)\alpha_{tj}+m\beta_t] 
\;.
\label{4.9}
\end{equation}
The multipliers $v_{s}$ are determined as the functions of $q^i$ and 
$p_i$ 
from the consistency for $\chi_{2s}$.
Hereafter we eliminate $\Pi_{s}$ and $\lambda^{s}$ with 
Eqs.(\ref{4.5}) and (\ref{4.9}), and regard $q^i$ and $p_i$ as the 
independent
 variables of the system.
The Poisson brackets of $\chi_{1s}$ and $\chi_{2s}$ are
\begin{equation}
\{\chi_{1s},\chi_{2t}\}=\frac{1}{m}E_{st} \;,
\label{4.10}
\end{equation}
which shows that all the constraints are second class.

Now the Dirac brackets for the variables become 
\begin{eqnarray}
\{q^i,q^j\}_{\rm D}&=&0\; ,
\label{4.11} \\
\{q^i,p_j\}_{\rm D}&=&\delta^{i}\!_{j}\; ,
\label{4.12} \\
\{p_i,p_j\}_{\rm D}&=&0\; .
\label{4.13}
\end{eqnarray}
The total Hamiltonian (\ref{4.6}) is reduced to the extended 
Hamiltonian
\begin{eqnarray}
H_{\rm E} & = & \frac{1}{2m}g^{ij}
[p_{i}-A_{i}-(E^{-1})^{st}\alpha_{si}
\{m\beta_{t}+g^{kl}(p_{k}-A_{k}) \alpha_{tl}\}]\nonumber\\
&& \mbox{}\times[p_{j}-A_{j}-(E^{-1})^{st}\alpha_{sj}
\{m\beta_{t}+g^{kl}
(p_{k}-A_{k})\alpha_{tl}\}]\nonumber\\
&& \mbox{}-\beta_{s}(E^{-1})^{st}[m\beta_{t}+g^{kl}(p_{k}-A_{k})
\alpha_{tl}]
+V(q)\; .
\label{4.14}
\end{eqnarray}
The classical dynamics of the system is described with this 
Hamiltonian and 
the Dirac brackets (\ref{4.11})--(\ref{4.13}).

In order to quantize the system, we must replace the Dirac brackets 
(\ref{4.11})--(\ref{4.13}) by the commutators ($\times 1/i\hbar$)
and regard 
variables $q^{i}$ and $p_{i}$ as operator variables $\hat{q}^{i}$ 
and 
$\hat{p}_{i}$. 
The results are  
\begin{eqnarray}
\left[\hat{q}^i,\hat{q}^j\right]&=&0\; , 
\label{4.15} \\
\left[\hat{q}^i,\hat{p}_j\right]&=&i\hbar\delta^i\!_j\; ,
\label{4.16} \\
\left[\hat{p}_i,\hat{p}_j\right]&=&0\; .
\label{4.17}
\end{eqnarray}
The commutation relations (\ref{4.15})--(\ref{4.17}) are canonical,
so that we easily obtain the coordinate representation of the 
momentum 
operators.
Symmetrizing the order of operators in Eq.(\ref{4.14}) 
to guarantee the hermiticity of the quantized Hamiltonian, we have
\begin{eqnarray}
\hat{H}_{\rm E}
& = & \frac{1}{8m}[\hat{F}^{k}\!_{i}\;,\;\hat{p}_{k}-\hat{A}_{k}]_{+}
\hat{g}^{ij}[\hat{F}^{l}\!_{j}\;,\;\hat{p}_{l}-\hat{A}_{l}]_{+} 
\nonumber \\
&& \mbox{}-\frac{1}{2}[\hat{g}^{ij}\hat{F}^{k}\!_{i}
(\hat{E}^{-1})^{uv}
\hat{\alpha}_{uj}\hat{\beta}_v\;,\;\hat{p}_{k}-\hat{A}_{k}]_{+} 
\nonumber \\
&& \mbox{}-\frac{1}{2}[(\hat{E}^{-1})^{uv}\hat{\beta}_{v}
\hat{\alpha}_{uj}\hat{g}^{ij}\;,\;\hat{p}_{i}-\hat{A}_{i}]_{+} 
\nonumber \\
&& \mbox{}+\frac{1}{2}m\hat{g}^{ij}(\hat{E}^{-1})^{uv}\hat{\beta}_{v}
\hat{\alpha}_{ui}(\hat{E}^{-1})^{st}\hat{\beta}_{t}\hat{\alpha}_{sj}
\nonumber \\
&& \mbox{}-m\hat{\beta}_{u}\hat{\beta}_{v}(\hat{E}^{-1})^{uv} 
\nonumber \\
&& \mbox{}+\hbar_{2}\hat{Q}+V(\hat{q})\; ,
\label{4.18}
\end{eqnarray}
where $\hat{F}^{j}\!_{i}$ is defined by
\begin{equation}
\hat{F}^{j}\!_{i}\equiv\left.\delta^{j}\!_{i}-\hat{g}^{jk}
(\hat{E}^{-1})^{st}\hat{\alpha}_{sk}\hat{\alpha}_{ti}\right|_{\hat{f}
=0}\; ,
\label{4.19}
\end{equation}
and $\hat{Q}$ is the quantum mechanical potential mentioned in 
Sec.~\ref{sec:3}.
 Henceforth we express all the operators in the coordinate
representation and omit the caret ``$\hat{}$''.

The requirement of hermiticity further leads us to the following two 
conditions:
\begin{equation}
F^{k}\!_{i}F^{l}\!_{j}\left(g^{ij}\left\{
\begin{array}{c}
a \\
ak
\end{array}
\right\}
-g^{jb}\left\{
\begin{array}{c}
i \\
kb
\end{array}
\right\}
\right)\partial_{l}+F^{l}\!_{i}g^{ij}(\partial_{k}F^{k}\!_{j})
\partial_{l}=0\;,
\label{4.20}
\end{equation}
\begin{equation}
(E^{-1})^{uv}\alpha_{uj}\beta_{v}g^{ij}(\delta^{k}\!_{i}+F^{k}\!_{i})
\left\{
\begin{array}{c}
l \\
lk
\end{array}
\right\}=0\; .
\label{4.21}
\end{equation}

The Schr\"{o}dinger equation of this system is written in the 
coordinate 
representation of the operators, with the conditions (\ref{4.20}) 
and 
(\ref{4.21}), as
\begin{eqnarray}    
i\hbar\frac{\partial \Psi}{\partial t}
&=& \hat{H}_{E}\Psi \nonumber \\
&=& \frac{1}{2m}g^{ij}F^{k}\!_{i}(-i\hbar \nabla_{k}-A_{k})
F^{l}\!_{j}(-i\hbar \nabla_{l}-A_{l})\Psi \nonumber \\
&& \mbox{} -g^{ij}(E^{-1})^{uv}\alpha_{uj}\beta_{v}F^{k}\!_{i}
(-i\hbar \nabla_{k}-A_{k})\Psi \nonumber \\
&& \mbox{} -\beta_{u}(E^{-1})^{uv}\alpha_{vl}g^{kl}
(-i\hbar \nabla_{k}-A_{k})
\Psi \nonumber \\
&& \mbox{} +\frac{1}{2}i\hbar[\nabla_k(E^{-1})^{uv}\alpha_{uj}
\beta_{v}g^{ij}
(\delta^{k}\!_{i}+F^{k}\,_{i})]\Psi \nonumber \\
&& \mbox{} +V(q)\Psi \;,
\label{4.22}
\end{eqnarray}
\[
\Psi\; : \;{\rm a \; wave \; function}\]
where $\nabla_{k}$ is the covariant derivative with respect to 
$q^{k}$, 
defined in Eq.(\ref{3.43}).

\section{SUMMARY AND DISCUSSIONS}

We have, in this paper, presented a general theory  of a particle 
constrained 
with several initial constraint conditions,
 thus aiming at completion of the dynamics of a particle on a 
 general manifold.

First we put the initial constraint conditions as functions of 
$q^{i}$.
 The Dirac brackets of $q^{i}$ and $p_{i}$ are complicated in 
 structure,
 which leads to some difficulties associated with quantization.
Therefore we apply the method of Homma {\it et al.} to the system 
and obtain 
the Schr\"{o}dinger equation.
In this process, requiring the Hamiltonian to be hermitian, we have 
the conditions (\ref{3.38})--(\ref{3.41}). 
{\it This is the result that was not obtained by Homma et al. 
and is due to the existence of several initial constraints in the 
system.} 

Next we treated the system of a particle constrained with
the initial constraint conditions (\ref{4.1}), depending not only on 
the coordinates but also on the velocities.  
The Dirac brackets of variables simply become canonical, so that 
we have the coordinate representation of the momentum operators
straightforwardly. 
We have the conditions (\ref{4.20}) and 
(\ref{4.21}) by requiring that 
the Hamiltonian should be hermitian.

Since $F^{j}\!_{i}$ depends on $f_{s}$ , we regard the hermiticity 
conditions 
as the conditions for the initial constraints $f_{s}$.
Hence, {\it in the quantum theory , the structure of the manifold
 on which the particle exists is to be restricted.}

Note that there is a case that the initial constraint conditions 
depend 
{\it explicitly} on time. 
In this case we cannot express the time evolution of physical 
quantities only with the Dirac bracket. We must make, at the 
same time, an explicit use of the Poisson bracket,
therefore we cannot apply the method of the present paper 
to this system.
The formalism of the quantization of this system is a subject for 
future study.

\bigskip

\noindent
{\large ACKNOWLEDGEMENTS}

The valuable discussions with S.Ishikawa are deeply appreciated.
One of the authors (M.Y) would like to thank Iwanami 
F${\rm\hat{u}}$jukai 
for financial support.  

\end{document}